\documentclass[aps,11pt,prd,longbibliography]{revtex4-1}
\usepackage{tikz}
\usepackage{feynmp-auto}
\usepackage{amsmath}
\usepackage[utf8]{inputenc}
\usepackage[T1]{fontenc} 
\usepackage{array}
\usepackage{bbm}
\usepackage{soul}	
\usepackage{xcolor}

\usepackage{epsfig}
\usepackage{color}
\usepackage{slashed}
\usepackage{comment}
\usepackage{epstopdf}
\usepackage{xspace}
\usepackage{mathrsfs}
\usepackage[euler]{textgreek}
\usepackage{amsmath}
\usepackage{amsthm}
\usepackage{amsfonts}
\usepackage{amssymb}
\usepackage{graphicx}
\graphicspath{ {Figures/} }
\usepackage[font={small}]{caption}
\usepackage{subcaption}
\usepackage{float}
\usepackage{multirow}
\usepackage[hang, flushmargin,bottom]{footmisc} 
\usepackage{placeins}
\usepackage{enumitem}
\usepackage{slashed}
\usepackage{bbm}
\usepackage{appendix}
\usepackage{tabularx}
\usepackage{siunitx}
\newcommand{\fb}[1]{\SI{#1}{\per\femto\barn}}
\newcommand{\ab}[1]{\SI{#1}{\per\atto\barn}}
\newcommand{\sqrts}[1]{$\sqrt{s}=\SI{#1}{TeV}$}

\usepackage{titlesec}
\titleformat{\chapter}[display]
  {\normalfont\LARGE\bfseries}
  {\chaptertitlename\ \thechapter}{5pt}{\LARGE}
  \titlespacing*{\chapter}{0pt}{-20pt}{35pt}
\usepackage{bigstrut}
\usepackage{pst-node}
\usepackage{pstricks}
\usepackage{physics}
\usepackage{mathtools}
\usepackage{setspace}
\usepackage{fancyhdr}
\usepackage[makeroom]{cancel}
\usepackage{feyn}
\newcommand{\be}{\begin{equation}}
\newcommand{\ee}{\end{equation}}
\newcommand{\bes}{\begin{equation*}}
\newcommand{\ees}{\end{equation*}}

\usepackage{hyperref}
\hypersetup{%
  colorlinks = true,
  linkcolor  = blue,
  citecolor  = blue,
  urlcolor   = blue,
}
\usepackage{soul}
\usepackage{todonotes}
\usepackage{xpatch}
\makeatletter
\xpretocmd{\todo}{\@bsphack}{}{}
\xapptocmd{\todo}{\@esphack}{}{}
\makeatother

\newcommand{\beq}{\begin{equation}}
\newcommand{\eeq}{\end{equation}}

\DeclareRobustCommand{\swatch}[1]{\tikz[baseline=-0.6ex]\node[fill=#1,shape=rectangle,draw=black,thick,minimum width=5mm,rounded corners=0.5pt](){};}

\newcommand{\MET}{\ensuremath{p_T^\mathrm{miss}}\xspace}
\newcommand{\gB}{\ensuremath{g_\mathrm{B}}\xspace}

\newcommand{\ZB}{\ensuremath{Z_\mathrm{B}}\xspace}
\newcommand{\HB}{\ensuremath{h_\mathrm{B}}\xspace}
\newcommand{\SB}{\ensuremath{S_\mathrm{B}}\xspace}
\newcommand{\MZB}{\ensuremath{M_{\ZB}}\xspace}
\newcommand{\MHB}{\ensuremath{M_{\HB}}\xspace}
\newcommand{\Mchi}{\ensuremath{M_{\chi}\xspace}}
\newcommand{\tB}{\ensuremath{\theta_\mathrm{B}}\xspace}
\newcommand{\stB}{\ensuremath{\sin\theta_\mathrm{B}}\xspace}

\newcommand{\GeV}{\text{GeV}}
\newcommand{\TeV}{\text{TeV}}

\newcommand{\herwig}{H\protect\scalebox{0.8}{ERWIG}\xspace}
\newcommand{\rivet}{R\protect\scalebox{0.8}{IVET}\xspace}

\newcommand{\hepdata}{HEPData\xspace}
\newcommand{\contur}{\textsc{Contur}\xspace}

\definecolor{green}{HTML}{008000}
\definecolor{goldenrod}{HTML}{DAA520}
\definecolor{magenta}{HTML}{FF00FF}
\definecolor{silver}{HTML}{C0C0C0}
\definecolor{indigo}{HTML}{4B0082}
\definecolor{skyblue}{HTML}{87CEEB}
\definecolor{darkgoldenrod}{HTML}{B8860B}
\definecolor{orange}{HTML}{FFA500}
\definecolor{yellow}{HTML}{FFFF00}
\definecolor{saddlebrown}{HTML}{8B4513}
\definecolor{blue}{HTML}{0000FF}
\definecolor{turquoise}{HTML}{40E0D0}
\definecolor{yellow}{HTML}{FFFF00}
\definecolor{white}{HTML}{FFFFFF}
\definecolor{whitesmoke}{HTML}{F5F5F5}

\newcommand{\myComment}[1]{}

\begin{document}
\title{\Large{Dark Matter from Anomaly Cancellation at the LHC}}
\author{Jon Butterworth$^{1}$, Hridoy Debnath$^{2}$, Pavel Fileviez P{\'e}rez$^{2}$, Yoran Yeh$^{1}$}
\affiliation{
$^{1}$Department of Physics and Astronomy, University College London, Gower St., London, WC1E 6BT, UK \\
$^{2}$Physics Department and Center for Education and Research in Cosmology and Astrophysics (CERCA), Case Western Reserve University, Cleveland, OH 44106, USA }
\email{j.butterworth@ucl.ac.uk, hxd253@case.edu,  pxf112@case.edu, yoran.yeh.20@ucl.ac.uk}
\date{\today}
\begin{abstract}
We discuss a class of theories that predict a fermionic dark matter candidate from gauge anomaly cancellation. 
As an explicit example, we study the predictions in theories where the global symmetry associated with baryon number is promoted to a local gauge symmetry.
In this context the symmetry-breaking scale has to be below the multi-TeV scale in order to be in agreement with the cosmological constraints on the dark matter
relic density. The new physical ``Cucuyo'' Higgs boson in the theory has very interesting properties, decaying mainly into two photons in the low mass region, and mainly into dark matter in the intermediate mass region. 
We study the most important signatures at the Large Hadron Collider, evaluating the experimental bounds. We discuss the correlation between the dark matter relic density, direct detection and collider constraints.
We find that these theories are still viable, and are susceptible to being probed in current, and future high-luminosity, running.
\end{abstract}
\maketitle
\section{INTRODUCTION}
Outstandingly successful though it is, the Standard Model (SM) of particle physics seems unlikely to be a final theory. The SM does not provide a mechanism for neutrino masses and an explanation for the matter-antimatter asymmetry of the universe. The SM also does not predict a candidate to describe the ``missing'' mass, or dark matter (DM), which could explain the rotation curves of galaxies and support the standard cosmological model. 

There is a long list of possible DM candidates, from  ultra-light fields to very heavy candidates such as primordial black holes~\cite{Chou:2022luk}. Candidates such as Axions and Weakly Interacting Massive Particles are still arguably the best motivated, due to the fact that they are predicted in several extensions of the SM which simultaneously solve other outstanding issues. As is well-known, a DM candidate should be electrically neutral and stable on the timescale of the age of the universe. One can look for DM signatures in several ways: a) in direct detection experiments through the scattering with matter, b) one can look for astrophysical signatures from DM annihilation, or c) look for missing energy signatures at colliders. While also taking the first possibility into account, we will study the third in detail a simple extension of the SM.

In this article, we study a class of theories where a DM candidate is predicted from the cancellation of gauge anomalies. If we consider an extension of the SM with extra gauge symmetries, one needs to make sure that the new gauge symmetry is anomaly free before investigating the predictions coming from spontaneous symmetry breaking. Instead of thinking about extra gauge symmetries not related to the SM, we investigate the case where the extra symmetry is just baryon number. In the SM baryon number is a global symmetry conserved only at the classical level but broken by $SU(2)_L$ instantons at the quantum level~\cite{tHooft:1976rip}. In the theories discussed in this article, the global baryon number of the SM is promoted to a local gauge symmetry. In this way one can study the implications from the spontaneous breaking of baryon number at the low energy scale.

In the simple theories proposed in Refs.~\cite{FileviezPerez:2011pt,Duerr:2013dza,FileviezPerez:2014lnj,FileviezPerez:2024fzc}, the spontaneous breaking of local baryon number is understood to be in agreement with existing experimental constraints. These theories predict a) a new fermionic sector needed for anomaly cancellation, b) the stability of the proton, because the interactions of the model violate baryon number by three units not by one unit, c) new sources of CP-violation and d) a fermionic DM candidate from anomaly cancellation. For several studies of the phenomenological and cosmological aspects of these theories see the studies in Refs.~\cite{Duerr:2013lka,Duerr:2014wra,Duerr:2017whl,FileviezPerez:2020oxn,FileviezPerez:2019jju,FileviezPerez:2018jmr}. In these studies it has been shown that using the cosmological bounds on the DM relic density, one can find an upper bound on the symmetry-breaking scale in the multi-TeV range. Therefore, one can hope to test these theories at the Large Hadron Collider (LHC) and its High Luminosity upgrade (HL-LHC).

In this article, we discuss the main signatures at the LHC and using the available experimental constraints we find the remaining allowed parameter space. We show that the new physical Higgs boson in the theory has very interesting properties because it decays mainly into two photons in the low mass region, and can decay mainly into dark matter in the intermediate mass region. We discuss the correlation between the cosmological bounds on the DM relic density, the bounds from direct detection experiments and the collider results obtained in this article. 
 
This article is organized as follows. In Section~\ref{Sec2} we discuss the main idea about the relation between DM and anomaly cancellation. In Section~\ref{Sec3} we discuss a simple theory for spontaneous local baryon number breaking where the predicted DM candidate is a Majorana fermion. In Section~\ref{Sec4} we discuss the main signatures at the LHC, and find out the allowed parameter space in the theory once use the experimental constraints from missing energy and dijet searches. In Section~\ref{Sec5} we discuss the correlation between the cosmological constraints on the relic density, the direct detection and collider constraints. In Section~\ref{Sec6} we discuss the future prospects to test this theory at the HL-LHC. We summarize the main results in Section~\ref{Sec7}.
\section{DARK MATTER FROM ANOMALY CANCELLATION}
\label{Sec2}
The SM is an anomaly-free quantum field theory that is very predictive. Anomaly cancellation is a powerful tool that imposes strong constraints on possible extensions of the SM. See Refs.~\cite{FileviezPerez:2010gw,FileviezPerez:2015mlm,Allanach:2018vjg,Ellis:2017tkh,FileviezPerez:2024fzc} for a recent discussion of anomaly cancellation.
In considering different theories for physics beyond the SM (BSM) where extra gauge symmetries exist, it is important to ensure that all new gauge anomalies are cancelled without spoiling the anomaly cancellation in the SM. Then, one can investigate the predictions from spontaneous symmetry breaking.

\begin{figure}[t]
\centering
\includegraphics[width=0.99\linewidth]{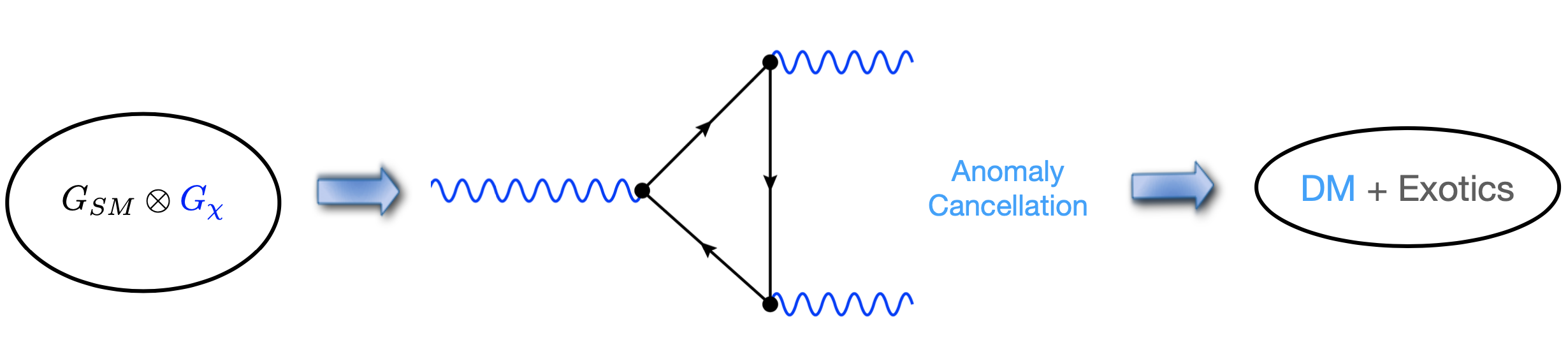}
\caption{Dark Matter from gauge anomaly cancellation. Here $G_{SM}$ is the SM gauge group, while $G_\chi$ is the new gauge group. The SM fields and the DM candidate are charged under the new gauge group. The new `exotic' fields together with the DM candidate are needed for anomaly cancellation.} 
\label{anomaly-DM}
\end{figure}

To simplify the discussion, let us focus on the case where the extra symmetry is an Abelian gauge group $U(1)^{'}$. The SM predicts two global symmetries associated with the conservation of baryon number (B) and total lepton number ($\ell$). These symmetries are good symmetries only at the classical level and are broken at the quantum level by $SU(2)_L$ instantons~\cite{tHooft:1976rip}. They play a major role in cosmology. One can define new gauge theories where either
or both, or combinations of them, are promoted to local gauge symmetries~\cite{FileviezPerez:2011pt,Duerr:2013dza,FileviezPerez:2014lnj,Amrith:2018yfb,FileviezPerez:2024fzc}. Since these symmetries are not anomaly free, one needs to add extra fermionic degrees of freedom to cancel the anomalies and then we can study the spontaneous breaking of these symmetries. If one of the extra fermionic degrees of freedom is neutral and automatically stable, the theory predicts a DM candidate from anomaly cancellation, as illustrated conceptually in Fig.~\ref{anomaly-DM}.

Let us now discuss in detail the predictions of a theory where the baryon number is a local gauge symmetry. That is, the gauge group $G_{\chi}=U(1)_B$. 
In this context, using the SM quark fields,
\begin{eqnarray}
q_L &\sim& ({\bf 3},{\bf 2},1/6,1/3), \ u_R \sim ({\bf 3},{\bf 1},2/3,1/3), \ {\rm{and}} \
d_R \sim  ({\bf 3},{\bf 1},-1/3,1/3), \nonumber
\end{eqnarray}
one can estimate the new gauge anomalies:
\begin{eqnarray*}
 {\mathcal{A}}(SU(3)_C^2 U(1)_B)&=&0, \
{\mathcal{A}}(SU(2)_L^2 U(1)_B)=3/2, \\
{\mathcal{A}}(U(1)_Y^2 U(1)_B)&=&-3/2, \
{\mathcal{A}}(U(1)_Y U(1)_B^2)=0, \\
{\mathcal{A}}(U(1)_B^3)&=&0, \ \text{and} \ 
{\mathcal{A}} (U(1)_B)=0.
\end{eqnarray*}
There are four simple ways to cancel the anomalies in these theories:
\begin{itemize}

\item Model I: In the simplest model proposed in Ref.~\cite{FileviezPerez:2024fzc}, one can cancel all anomalies with only four extra fermionic representations.

\item Model II: Adding six extra colorless fermionic representations~\cite{Duerr:2013dza}.

\item Model III: Adding four extra colorless fermionic representations~\cite{FileviezPerez:2014lnj}, where there of the representations transform non-trivially under $SU(2)_L$.

\item Model IV: Adding vector-like quarks~\cite{FileviezPerez:2011pt}. 

\end{itemize}

In the first three cases, the simplest scenarios, one always predicts a DM candidate from anomaly cancellation. In the second case the field, $\chi_L \sim (1,1,0,n_\chi)$, can be a Majorana or Dirac field, while in the first and third cases, it has to be Majorana. 
For a more detailed discussion of these theories see for example Refs.~\cite{FileviezPerez:2015mlm,Ellis:2017tkh,FileviezPerez:2024fzc}. In the next section, we will discuss the signatures associated with the DM candidate in the models where one predicts such a candidate from anomaly cancellation. We will discuss the scenarios in which the DM is a Majorana fermion, since this is the most generic case that can be realized in models I, II and III mentioned above.
\section{DARK MATTER AND LOCAL BARYON NUMBER}
\label{Sec3}
As mentioned above, one predicts a fermionic DM candidate from gauge anomaly cancellation in a theory where the baryon number is a local symmetry. The baryon number can be defined as a local gauge symmetry and one can have a simple theory based on $$SU(3)_C \otimes SU(2)_L \otimes U(1)_Y \otimes U(1)_B,$$
as proposed in Refs.~\cite{Duerr:2013dza,FileviezPerez:2014lnj,FileviezPerez:2024fzc}.
The baryon number is not anomaly-free and one has to add extra fermions to cancel all anomalies. 
As discussed above, there are four simple realistic models for anomaly cancellation, for more details see Refs.~\cite{FileviezPerez:2011pt,Duerr:2013dza,FileviezPerez:2014lnj,Duerr:2013lka,Duerr:2014wra,Ohmer:2015lxa,FileviezPerez:2024fzc}. 
These theories predict a new leptophobic gauge boson, \ZB, a new Higgs, \SB, and extra fermions.
Additionally, in the theories studied in Refs.~\cite{Duerr:2013dza,FileviezPerez:2014lnj,FileviezPerez:2024fzc}, one predicts a DM candidate, $\chi$, from anomaly cancellation.

In these theories one has the following interactions:
\begin{eqnarray}
{\cal L} &\supset& -\frac{\gB}{3} \bar{q} \gamma^\mu q Z_\mu^B \ + \ (D_\mu \SB)^\dagger (D^\mu \SB) 
+ i \bar{\chi}_L \gamma^\mu D_\mu \chi_L - V(H,\SB) - \left( \lambda_\chi \chi_L^T C \chi_L \SB \ + \  {\rm{h.c.}} \right) \nonumber \\
&+& {\cal L}_{\rm{new}},
\end{eqnarray}
with $\SB \sim (1,1,0,n_B)$, $\chi_L \sim (1,1,0,n_\chi)$, and 
\begin{eqnarray}
D^\mu \SB &=& \partial^\mu \SB + i \gB n_B Z^\mu_B \SB, \\
D^\mu \chi_L &=& \partial^\mu \chi_L + i \gB n_\chi Z^\mu_B \chi_L,
\end{eqnarray}
and ${\cal L}_{\rm{new}}$ contains the extra fields needed for anomaly cancellation. In these theories $n_B=3$ and $n_\chi=-3/2$ in the Majorana case. Notice that this theory predicts that the proton is stable because the interactions of the model break local baryon number by three units. In this article we stick to the scenarios where the fermionic DM candidate is a Majorana fermion because this scenario can be realized in both models proposed in Refs.~\cite{Duerr:2013dza,FileviezPerez:2014lnj,FileviezPerez:2024fzc}.

The scalar potential of this theory reads as
\begin{eqnarray}
	V &=& -\mu_H^2 H^{\dagger} H + \lambda_H (H^{\dagger} H)^2 - \mu_B^2 S^{*}_B \SB + \lambda_B (S^{*}_B \SB)^2  
 + \lambda_{HB} (H^{\dagger}H)(S^{*}_B \SB),
	\label{eq:potential}
\end{eqnarray}
and the Higgs bosons can be written as
\begin{align}
	H = \mqty(h^+ \\ \frac{1}{\sqrt{2}}(h_0 + i a_0)),\quad  \text{ and } \quad \SB = \frac{1}{\sqrt{2}}(s_B + i a_B).
\end{align}
Here $a_0$ and $a_B$ are the Nambu-Goldstone's bosons.
Spontaneous symmetry breaking of baryon number is achieved once 
\SB acquires a vacuum expectation value (vev), $\langle s_B \rangle = v_B$, while the electroweak spontaneous symmetry is triggered by the vev of the SM Higgs, $\langle h_0 \rangle = v_0$. The Higgs bosons mix through the scalar potential in Eq.~\eqref{eq:potential}. In the broken phase, 
the physical Higgs bosons are defined as 
\begin{equation}
\begin{split}
	h &= h_0 \cos{\tB} - s_B \sin{\tB},\\
	\HB &= s_B \cos{\tB} + h_0 \sin{\tB},
	\end{split}
\end{equation}
where the mixing angle $\tB$ that diagonalizes the mass matrix for the Higgs bosons is given by
\begin{align}
	\tan{2 \tB} = \frac{v_0 v_B \lambda_{HB}}{ v_B^2 \lambda_B - v_0^2 \lambda_H }.
\end{align}
In the theories proposed in Refs.~\cite{Duerr:2013dza,FileviezPerez:2014lnj,FileviezPerez:2024fzc} the anomalies cancel in different ways. 
These fields can be light, but here we will assume for simplicity that they are close to the upper bound on the symmetry breaking scale, i.e. at the multi-TeV scale. For previous studies see Refs.~\cite{FileviezPerez:2020oxn,FileviezPerez:2021tgo,FileviezPerez:2020mta,Duerr:2017whl,FileviezPerez:2019jju}.
In Ref.~\cite{FileviezPerez:2018jmr} the need to impose perturbative bounds on the scalar couplings in the Higgs sector was pointed out:
\begin{align}
\lambda_H = & \frac{1}{2v_0^2} \left( M_{h}^2 \cos^2 \tB + \MHB^2 \sin^2 \tB \right) \leq 4 \pi, \\
\lambda_{B} = & \frac{1}{2v_{B}^2} \left( M_{h}^2 \sin^2 \tB + \MHB^2 \cos^2 \tB \right) \leq 4 \pi, \\
\lambda_{HB} = & \frac{1}{v_0 v_{B}} \left( \MHB^2  - M_{h}^2 \right) \sin \tB \cos \tB  \leq 4 \pi .
\end{align}
To ensure vacuum stability of the scalar potential we impose
\begin{align}
\lambda_H, \lambda_{B} > 0 \hspace{5mm} {\rm and} \hspace{5mm} \lambda_H \lambda_B - \lambda_{HB}^2/4 > 0,
\end{align}
for more details see Ref.~\cite{FileviezPerez:2018jmr}. 
For our calculation to remain consistent, the gauge coupling \gB must remain perturbative. If we revisit all the gauge interactions present in the theory, symbolically we have $\bar{q} \ZB q$, $\bar{\chi} \ZB \chi$, and $\ZB \ZB \SB \SB$, and
we find that the strongest upper bound on \gB comes from the $\ZB \ZB \SB \SB$ interaction. It reads as $\gB \leq \sqrt{2 \pi}/3$. Following the notation above, the perturbative bound on the Yukawa coupling between the DM and the Higgs bosons is $\lambda_\chi < 2 \sqrt{\pi}$.

After the local gauge symmetry is broken one has the ${\mathcal{Z}}_2$ symmetry: $$\chi \to - \chi,$$ protecting the DM candidate stability. Here $$\chi=\chi_L + (\chi_L)^C,$$ is the Majorana DM field.
It is important to mention that the stability of the DM candidate is a natural consequence of the spontaneous symmetry breaking. 
Before discussing some phenomenological aspects, we emphasize that these theories have only a few relevant parameters:
$$n_\chi, M_\chi, \gB, \MZB, \MHB, \ {\rm{and}} \ \tB.$$
In our study we will use $n_\chi=-3/2$ as predicted in the Majorana case in the theories discussed above. 
\subsection{Leptophobic Gauge Boson}
\begin{figure}[h]
\centering
\includegraphics[width=0.55\linewidth]{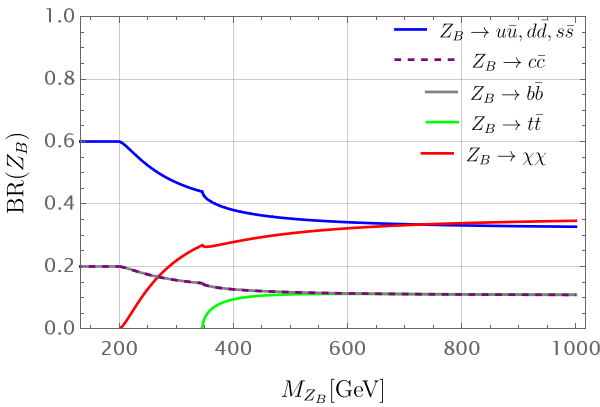}
\caption{Branching ratios of \ZB decays as a function of \MZB when $M_{\chi}=100 $ \GeV. The branching ratios are independent of \MHB, $\gB$ and \stB.
} 
\label{ZBBRs}
\end{figure}

These theories predict a new massive gauge boson, $\ZB$, that couples mainly to quarks, the DM candidate and the new fermions needed for anomaly cancellation with baryon number. In Fig.~\ref{ZBBRs} we show the branching ratios for the \ZB decays to understand the importance of the invisible decays. Here we assume for simplicity that the DM mass is $M_\chi=100$ GeV, and show the decays into light quarks (in blue), the decays into charm quarks (dashed-line), decays into bottom quarks (in gray), and the decays into top quarks (in green).
When the invisible decays are allowed the branching ratio can be as high as $35\%$. The decays into two top quarks have a branching ratios only around $10\%$, but they are very important to identify this gauge boson at the LHC due to fact one can trigger on and reconstruct the top-pair much better above the QCD background, compared to light jet production.

\begin{figure}[h]
\centering
\includegraphics[width=0.9\linewidth]{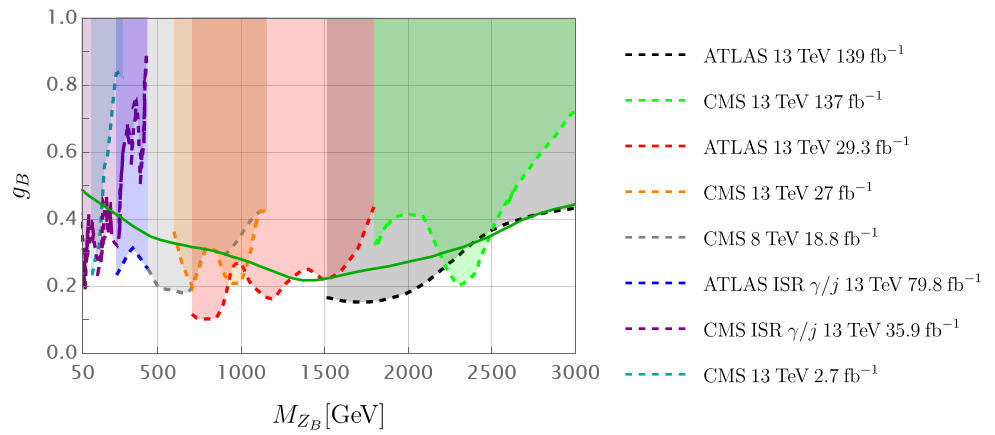}
\caption{Experimental bounds on the leptophobic gauge boson \ZB mass~\cite{2020,Aaboud_2018,Aaboud_2019,Sirunyan_2020,Khachatryan_2016,Sirunyan_2017,2018,Sirunyan_2019e}, assuming \HB and $\chi$ are decoupled.
The green line is the bound from \contur taken from Fig.~\ref{fig:conturScan10}, which does not assume decoupling (see text).
} 
\label{ZBbounds}
\end{figure}

Before studying all the signatures at the LHC, we examine the current collider bounds on a generic leptophobic \ZB as a function its mass. In Fig.~\ref{ZBbounds} we show the current collider bounds from the CMS and ATLAS collaboration for a \ZB which decays only to quarks, which in our model corresponds to a decoupled exotic Higgs sector and suppressed $\ZB \to \chi\chi$ decays. The bounds shown in Fig.~\ref{ZBbounds} are as follows:
\begin{itemize}
\item The black dashed line shows the ATLAS bounds at \sqrts{13} with \fb{139} integrated luminosity.  In this search they look for new resonances decaying into a pair of jets~\cite{2020}.

\item The green dashed line shows CMS bounds when \sqrts{13} with \fb{137}. In this case a search for narrow and broad resonances with masses greater than $1.8$ TeV decaying to a pair of jets was performed~\cite{Sirunyan_2020}.

\item The red dashed line shows the ATLAS bounds when \sqrts{13} and \fb{29.3}. This search targets low-mass dijet resonances in the range $450-1800$ GeV~\cite{Aaboud_2018}.

\item The orange dashed line shows the CMS bounds when \sqrts{13} TeV and \fb{27} ~\cite{2018}. Here they look for searches for resonances decaying into pairs of jets.

\item The gray dashed line shows the CMS bounds with \sqrts{8} with \fb{18.8}. In this case a search for narrow resonances decaying into dijet final states was performed and the data were collected with the CMS detector using a novel technique called data scouting, in which the information associated with these selected events is much reduced, permitting collection of larger data samples~\cite{Khachatryan_2016}.

\item The blue dashed line shows the ATLAS bounds when using the $\gamma/j$ initial state radiation with \sqrts{13} and \fb{79.8}. In this case a search was performed for localised excesses in dijet mass distributions of low-dijet-mass events produced in association with a high transverse energy photon~\cite{Aaboud_2019}.

\item The purple dashed line shows the CMS bounds when one uses $\gamma/j$ initial state radiation with \sqrts{13} with \fb{35.9} (2016) and \fb{41.1} (2017)~\cite{Sirunyan_2019e}.
\end{itemize}
The green solid line shows, for comparison, the \contur bounds discussed in the next section. As one can see, the leptophobic gauge boson \ZB can be light, in fact its mass can be very close to the electroweak scale without implying a very small gauge coupling \gB.  
%
\subsection{The Cucuyo-Higgs Decays}
These theories predict the existence of a second physical Higgs boson, \HB.
\begin{figure}
         \centering
    \begin{subfigure}[b]{0.45\textwidth}
         \centering
         \includegraphics[width=\textwidth]{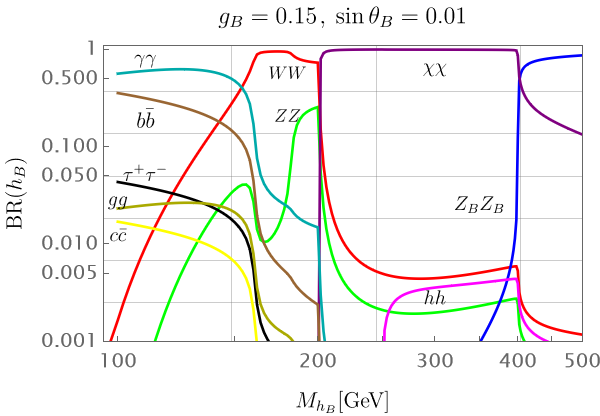}
         \caption{}
    \end{subfigure}  
         \begin{subfigure}[b]{0.45\textwidth}
         \centering
         \includegraphics[width=\textwidth]{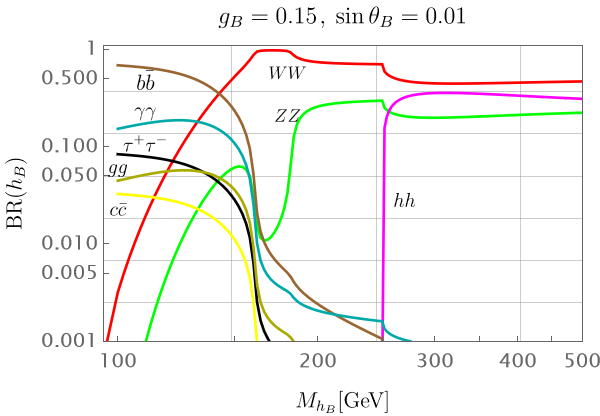}
         \caption{}
    \end{subfigure} 
        \caption{(a) Branching ratios of \HB decays when $M_{\chi}= 100$ GeV and $\MZB=200$ GeV. (b) as in (a), but now with $M_{\chi}= 300$ GeV and $\MZB=500$ GeV. Here we assume $\gB=0.15$, $\sin \tB=0.01$, and masses of the new charged fermions are 1 TeV in both cases.
}
        \label{HiggsDecay}
\end{figure}
The main Higgs decays are $\HB \to \gamma \gamma, WW, ZZ, b\bar{b}, hh, \ZB\ZB, \chi \chi$, while the other decays have small branching ratios. The predictions of these decays depend on the mixing angle, \tB, between the SM Higgs and \HB, and of course of the value of the new gauge coupling, \gB. (See the appendix for the Feynman rules.) As we will discuss in the next section, the mixing angle has to be small in order to satisfy the experimental bounds from direct-detection DM experiments.

In Fig.~\ref{HiggsDecay} we show the predictions for the \HB branching ratios in two scenarios. In the left panel we assume $\gB=0.15$ and $\sin \tB=0.01$. Decays of \HB in the low mass region are different from SM-like Higgs. Notice that the branching ratio into two photons can be very large, and for this reason we refer to the \HB as the {\textit{Cucuyo}}~\footnote{An insect very common in Cuba, Brazil, Guiana and Mexico. It may be seen at night in great numbers amongst the foliage of trees. They sometimes are so numerous that they light up the dark forest.}-Higgs. Here the effective coupling between \HB and the two photons is generated at one-loop level where inside the loop we have the new electrically-charged fermions needed for anomaly cancellation. Here we used the fields in the minimal model proposed in Ref.~\cite{FileviezPerez:2014lnj} where only four extra fermionic representations are needed. See also the study in Ref.~\cite{FileviezPerez:2020oxn} for the Feynman rules in this model. 

One can see in Fig.~\ref{HiggsDecay} that in the high-mass region the decays to two \ZB and to DM can be important. Here we use the same value for the mixing angle, $\sin \tB=0.01$. In the intermediate mass region decays into the DM candidate can dominate. However, when the decays into two \ZB gauge bosons are allowed they can have very large branching ratios. 
It is important to mention that the experimental bounds on the new Higgs mass are very weak because one cannot produce this new Higgs with very large cross sections.
The production mechanism $pp\to \ZB^* \to \ZB \HB$ is unique in this theory. Although this cross section is not very large, one can have unique signatures when the \HB decays mainly into two photons and \ZB into two top quarks, i.e. $pp\to \ZB^* \to \ZB \HB \to t \bar{t} \gamma \gamma$. We will discuss all possible signatures in the next section and show the allowed regions of the parameter space.
%
\section{LHC CONSTRAINTS AND SIGNATURES}
\label{Sec4}
%
\begin{figure}
\centering
    
    \begin{subfigure}[a]{0.9\textwidth}
         \centering
         \includegraphics[width=\textwidth]{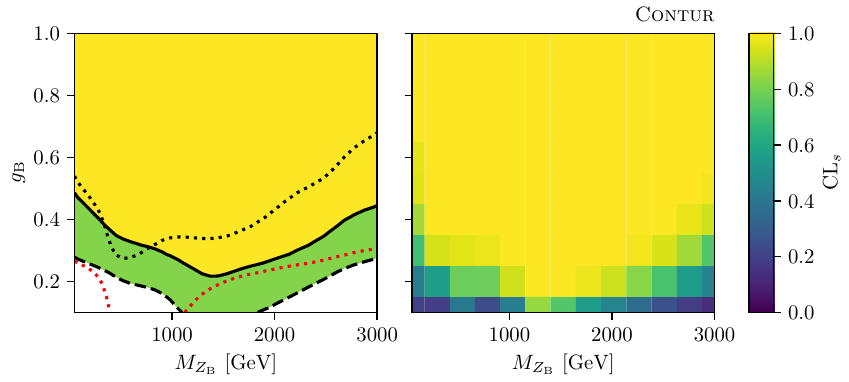}
         \caption{}
    \end{subfigure}  
    \begin{subfigure}[b]{0.45\textwidth}
         \centering
         \includegraphics[width=\textwidth]{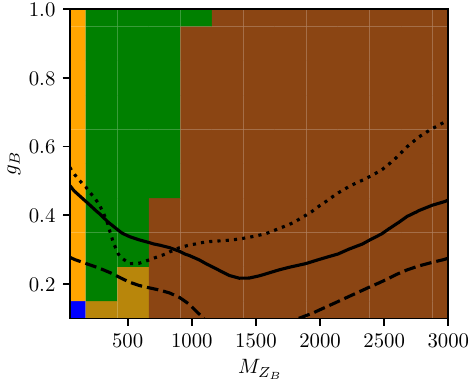}
         \caption{}
    \end{subfigure}
    \begin{subfigure}[c]{0.45\textwidth}
\vspace{-6cm}
        \begin{tabular}{ll}
        \swatch{green}~\MET{}+jet \cite{ATLAS:2017txd,ATLAS:2024vqf} \\
        \swatch{orange}~$\ell^+\ell^-$+jet \cite{ATLAS:2015iiu} \\ 
        \swatch{saddlebrown}~hadronic $t\bar{t}$ \cite{ATLAS:2022mlu,ATLAS:2020ccu} \\ 
        \swatch{darkgoldenrod}~$\ell^+\ell^-\gamma$ \cite{ATLAS:2019gey} \\
        \swatch{blue}~$\ell$+\MET{}+jet \cite{ATLAS:2017irc,ATLAS:2015mip,ATLAS:2017luz} 
        \end{tabular}
    \end{subfigure}
        \caption{Results from \contur, showing the 95\% exclusion (solid line), the 68\% exclusion (dashed line) and the expected 95\% exclusion (dotted line). In a) the total exclusion is shown, indicating the granularity of the scan
        on which the interpolated exclusions are based. In this plot the red dotted line
        also gives an indication of the eventual sensitivity of the HL-LHC (see text). In b), the contours are the same but the colouring indicates which signature gives the greatest exclusion at each point.
        The fixed parameters for this scan are \MHB = \SI{200}{GeV}, $\stB=0.01$, and $M_\chi=\SI{100}{GeV}$. 
        Citations are for the most significant measurements giving an exclusion significance of 68\% or above anywhere in the plain.}
        \label{fig:conturScan10}
\end{figure}
The collider phenomenology of the theory discussed above includes several interesting signatures related to the leptophobic gauge boson, \ZB,
the Majorana DM candidate, $\chi$, and the new Higgs boson, \HB.
The coupling of \ZB to quarks implies dijet production $q \bar{q} \to \ZB^* \to \mathrm{jets}$, and for sufficiently high \MZB, $q \bar{q} \to \ZB^* \to t \bar{t}$.
Both can be visible as resonant or non-resonant production depending on \MZB. If the decay $\ZB \to \chi\chi$ is kinematically allowed, missing
energy signatures will be present from associated \ZB production:
$q \bar{q} \to g \ZB^* \to g \chi\chi,  \ g q \to q \ZB^* \to q \chi\chi, \ \ g \bar{q} \to \bar{q} \ZB^* \to \bar{q} \chi\chi$. At lower cross sections,
but with potentially more distinctive signatures, associated production of \ZB with a gauge boson will also be present,
for example $q \bar{q} \to \gamma \ZB^* \to \gamma \chi\chi$.
There are also potential signatures from associated production of \HB; $q \bar{q} \to \ZB^* \to \ZB \HB$, with subsequent decay of \ZB giving rise to
jets or \MET, and the \HB decays depending on its mass as shown in Fig.~\ref{HiggsDecay}.

Other possibilities involving the SM Higgs boson exist, in which the production cross section is proportional to the mixing angle between the two Higgs bosons.
However, we will focus on those possibilities where the production mechanisms is not suppressed by the mixing angle, as the angle is already constrained
to be fairly small by SM Higgs measurements, and the above signatures are more generic in the sense that they do not depend on it strongly. Notice that the mixing angle between the Higgses has to very small in order to satisfy the DM direct detection bounds, see the discussion in the next section.

Several measurements of relevant final states have been made at the LHC and shown to be consistent with SM predictions. Many of these measurements
are available in \hepdata~\cite{Maguire:2017ypu} and in the \rivet~\cite{Bierlich:2019rhm} analysis library,
and we will use the \contur~\cite{Butterworth:2016sqg,Buckley:2021neu} analysis tool to evaluate the constraints they place on the parameters of the model.
For a given point in model parameter space, 
events are simulated using our Feynrules~\cite{Alloul:2013bka} implementation of the theory, exported
as UFO~\cite{Degrande:2011ua} files and read by the \herwig~\cite{Bahr:2008pv,Bellm:2015jjp,Bewick:2023tfi} event generator.
\contur performs a signal injection of the \herwig events into the fiducial phase space of the measurements.
Hundreds of such measurements are available, and thus a wide range of possible signatures can be probed simultaneously.
The level of agreement between data and the SM is evaluated, as is the agreement between SM+BSM and the data.
The likelihood ratio of the two, calculated using the $CL_S$ method~\cite{Read:2002hq} is used to determine an exclusion limit; the expected exclusion limit is also determined by setting the data to the SM values but preserving its uncertainties.
Since all the measurements used are at least consistent with the SM, exclusions are expected. However, if small discrepancies accumulate,
the first hint of the presence of BSM physics in the data could be a divergence between the expected and the actual exclusion.
This method was previously used to study a leptophobic $Z^\prime$ in topcolour models, showing that measurements of top quark and jet production do have sensitivity~\cite{Altakach:2021lkq}.

As a starting point, following Fig.~\ref{ZBbounds}, we perform a course-grained parameter scan over a wide range of \gB ($0.1 < \gB < 1.0$) and \MZB ($50~\GeV < \MZB < 3000~\GeV$), for $M_\chi = 100~\GeV$, $\stB=0.01$ and a Cucuyo Higgs mass $\MHB=200$~GeV.
The result is shown in Fig.~\ref{fig:conturScan10}. In the right hand portion of 
Fig.~\ref{fig:conturScan10}a, the granularity of the scan can be seen, with the exclusion for each point indicated by the colour bar. The left hand portion of the same plot shows the interpolated exclusion contours derived from this, along with
the expected exclusion. The model is excluded over much of the plain considered, above $\gB \approx 0.3$. The \contur exclusion is superimposed on Fig.~\ref{ZBbounds} for ease of comparison; however, it should be borne in mind that, on the one hand some of 
the data used in the \ZB searches is not yet used in measurements available in \contur, and on the other hand, the \contur limits are able to consider the exact model parameters, including potential $\chi$ and \HB production, which are not considered in the \ZB searches, and which can, for example, reduce the resonant dijet or $t\bar{t}$ signal in cases where the $\ZB \to \chi\chi$ branching fraction is significant.
 
Fig.~\ref{fig:conturScan10}b shows the same contours, but now superimposed on a colour scheme which indicates which final-state signature is most responsible
for the exclusion at each point.
At high \MZB, the strongest sensitivity comes from fully-hadronic top measurements, with the signature being high-$p_T$, boosted top quarks from \ZB decays.
As \MZB approaches the threshold for top pair production, the relative importance shifts in favour of \MET signatures, which arise from $\ZB \to \chi\chi$ decays
recoiling against jets.
At even lower \MZB the sensitivity is principally due to $Z\ZB$ associated production, where the \ZB decays to jets and the $Z$ decays to leptons,
and comes from precision $Z+$jets measurements, made with \sqrts{8} LHC data.
Note that all signatures are studied for each point, so typically several final states contribute to the exclusion, while the colour scheme only
indicates which gives the largest contribution.
\begin{figure}
         \centering
    \begin{subfigure}[b]{0.48\textwidth}
         \centering
         \includegraphics[width=\textwidth]{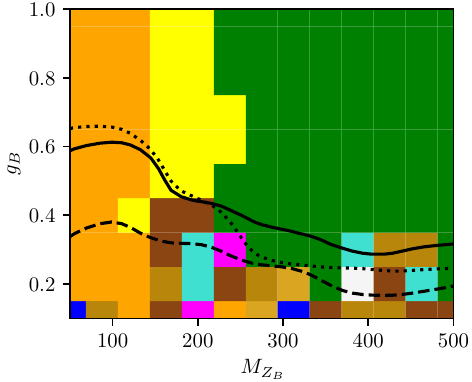}
         \caption{}
         \label{fig:conturScan11a}
    \end{subfigure}  
         \begin{subfigure}[b]{0.48\textwidth}
         \centering
         \includegraphics[width=\textwidth]{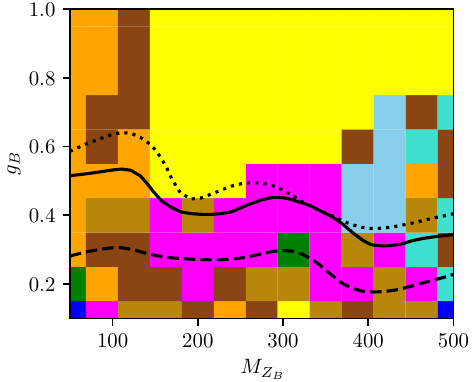}
         \caption{}
         \label{fig:conturScan11b}
    \end{subfigure} 
\vspace{0.5cm}
    \begin{tabular}{llll}
        \swatch{turquoise}~$\ell_1\ell_2$+\MET{}+jet \cite{ATLAS:2019ebv,ATLAS:2021jgw} & 
        \swatch{yellow}~$\gamma$ \cite{ATLAS:2021mbt,ATLAS:2017xqp} & 
        \swatch{goldenrod}~$\gamma$+\MET{} \cite{ATLAS:2018nci} & 
        \swatch{silver}~jets  \\ 
        \swatch{green}~\MET{}+jet \cite{ATLAS:2017txd,ATLAS:2024vqf} & 
        \swatch{darkgoldenrod}~$\ell^+\ell^-\gamma$  & 
        \swatch{saddlebrown}~hadronic $t\bar{t}$ \cite{ATLAS:2022mlu,ATLAS:2020ccu} & 
        \swatch{whitesmoke}~$\ell^\pm\ell^\pm$+\MET{} \cite{ATLAS:2019cbr} \\
        \swatch{orange}~$\ell^+\ell^-$+jet \cite{ATLAS:2020juj,ATLAS:2019ebv} &
        \swatch{skyblue}~$\ell_1\ell_2$+\MET{} \cite{ATLAS:2019rob} & 
        \swatch{magenta}~4$\ell$ \cite{ATLAS:2019qet,ATLAS:2021kog} & 
        \swatch{blue}~$\ell$+\MET{}+jet \cite{ATLAS:2017luz}\\ 
    \end{tabular} 
        \caption{(a) Results from \contur, as in Fig.~\ref{fig:conturScan10}b, but now zooming in
        on the low \MZB region. (b) as in (a), but now with $M_\chi=10\,\text{TeV}$. Citations are for the most significant measurements giving an exclusion significance of 68\% or above anywhere in the plain.}
        \label{fig:conturScan11}
\end{figure}

To study the low \MZB region in more detail, a higher-granularity scan was performed, shown in Fig.~\ref{fig:conturScan11}a.
Here the \MET signatures dominate the sensitivity where $\MZB > 2M_\chi$, as expected from the coarse scan. At lower \MZB, this signature gradually switches off,
and the remaining sensitivity comes from isolated photon measurements and the \sqrts{8} precision $Z+$jets measurement.
The \MET signature can be turned off completely if $M_\chi \gg \MZB/2$, as shown in Fig.~\ref{fig:conturScan11}b. In this region
\MZB is too small for $\ZB \to t\bar{t}$ to feature, and $\ZB$ decays to light quarks suffer from high SM backgrounds. However, 
sensitivity remains in the prompt photon measurements due to 
$p p \to \ZB \HB$ production with the Cucuyo Higgs decays, $\HB \to \gamma\gamma$, lighting up the dark sector.
As mentioned above, this decay is mediated by loops which can involve the new fermions of the model needed for anomaly cancellation. For the details about the new fermions in the minimal model see Refs.~\cite{FileviezPerez:2014lnj,FileviezPerez:2020oxn}.
Note that the $\HB \rightarrow ZZ$ and $\HB \to WW$ branchings are also relatively large, and both make sub-dominant contribution to the exclusion in the leptonic decay channels. At low \MZB, again the $Z+$jets measurement contributes; there is also a contribution from boosted hadronic top measurements, coming from off-shell $q\bar{q} \to \ZB^* \to t\bar{t}$ processes.

\begin{figure}
    \begin{subfigure}[b]{0.45\textwidth}
        \centering
        \includegraphics[width=\textwidth]{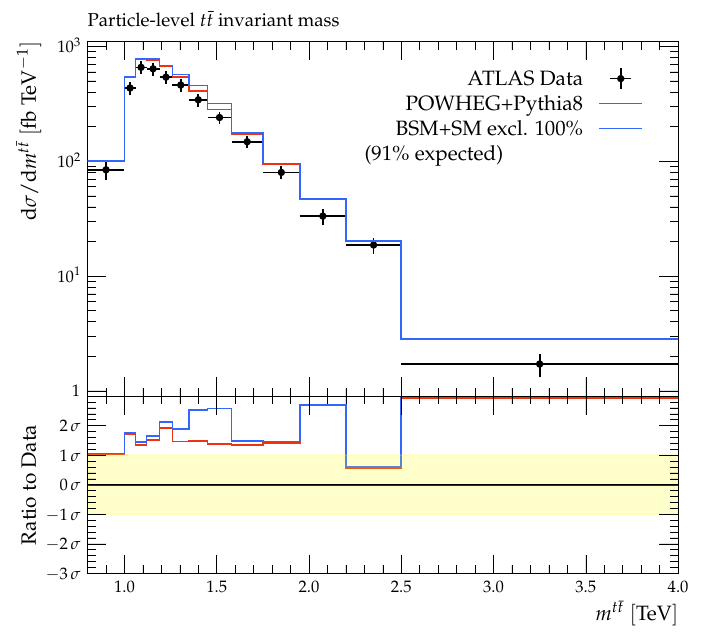}
        \caption{}
    \end{subfigure}
         \centering
    \begin{subfigure}[b]{0.45\textwidth}
         \centering
         \includegraphics[width=\textwidth]{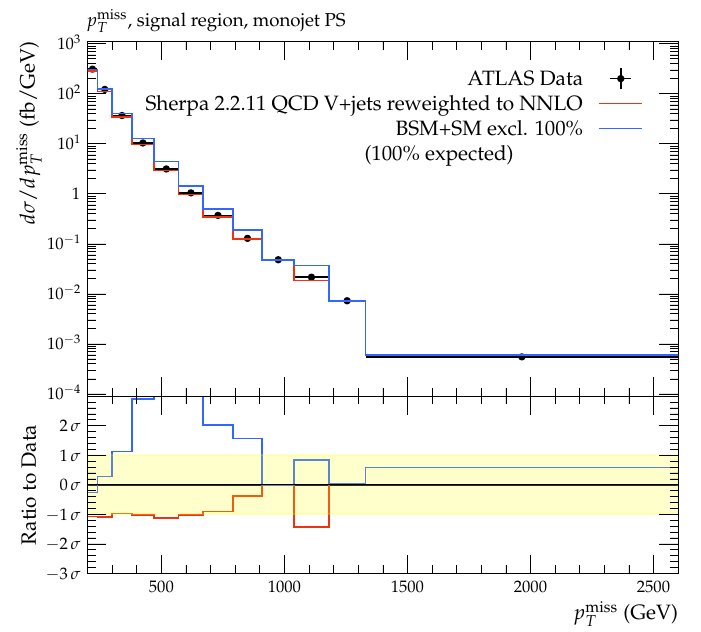}
         \caption{}
    \end{subfigure}  
         \begin{subfigure}[b]{0.45\textwidth}
         \centering
         \includegraphics[width=\textwidth]{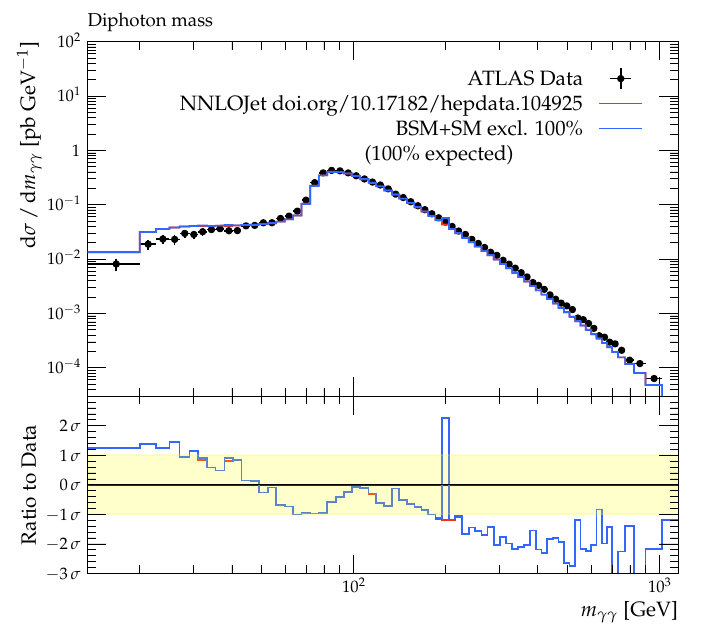}
         \caption{}
    \end{subfigure} 
        \caption{Subset of differential histograms that contribute to collider constraints from experimental measurements. (a) Hadronic $t\bar{t}$ measurement \cite{ATLAS:2022mlu}, corresponding to Fig.~\ref{fig:conturScan10} for $\MZB=1525\,\text{GeV}$ and $\gB=0.4$. The
        SM prediction using POWHEG~\cite{Alioli:2010xd} is taken from the experimental paper.
        (b) Missing energy in association with jets measurement \cite{ATLAS:2024vqf}, corresponding to Fig.~\ref{fig:conturScan11a} for $\MZB=425\,\text{GeV}$ and $\gB=0.4$. The
        SM prediction using Sherpa~\cite{Sherpa:2019gpd} is taken from the experimental paper. (c) Di-photon measurement \cite{ATLAS:2021mbt}, corresponding to Fig.~\ref{fig:conturScan11b} for 
        $\MZB=254\,\text{GeV}$ and $\gB=0.6$. The SM prediction using NNLOJet~\cite{Gehrmann:2018szu} is taken from the experimental paper.}
        \label{fig:RivetPlots}
\end{figure}

\begin{figure}
     \centering
    \begin{subfigure}[b]{0.48\textwidth}
         \centering
         \includegraphics[width=\textwidth]{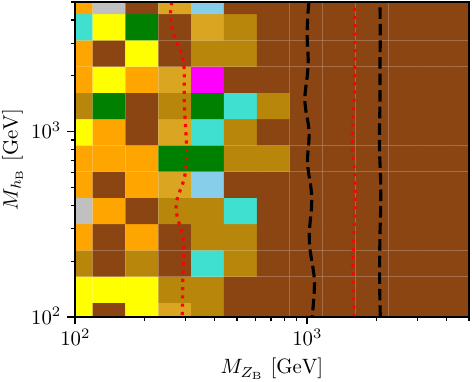}
         \caption{}
    \end{subfigure}  
\begin{subfigure}[b]{0.48\textwidth}
         \centering
         \includegraphics[width=\textwidth]{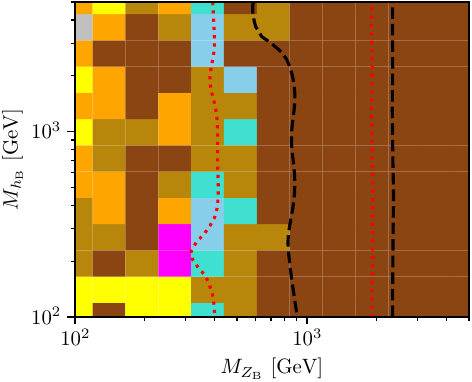}
         \caption{}
    \end{subfigure}  
 \begin{subfigure}[b]{0.48\textwidth}
         \centering
         \includegraphics[width=\textwidth]{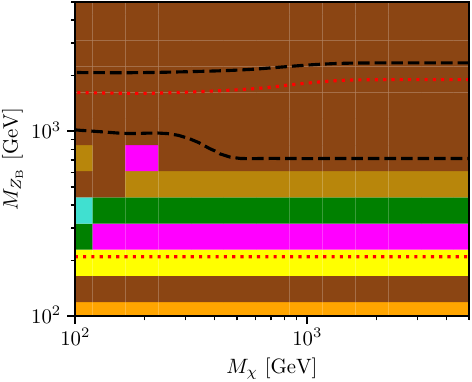}
         \caption{}
     \end{subfigure}
      \begin{subfigure}[b]{0.48\textwidth}
         \centering
         \includegraphics[width=\textwidth]{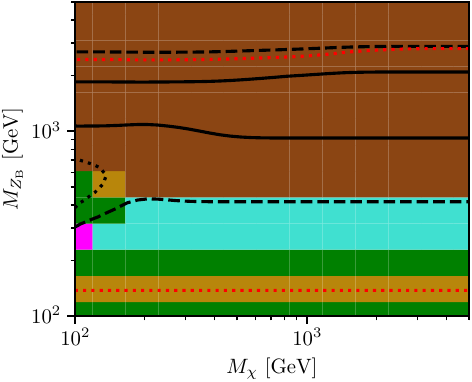}
         \caption{}
     \end{subfigure}
%
    \begin{tabular}{llll}
        \swatch{yellow}~$\gamma$ \cite{ATLAS:2014yga,ATLAS:2021mbt} &
        \swatch{goldenrod}~$\gamma$+\MET{} \cite{ATLAS:2016qjc,ATLAS:2018nci} &
        \swatch{darkgoldenrod}~$\ell^+\ell^-\gamma$ \cite{ATLAS:2019gey} & 
        \swatch{saddlebrown}~hadronic $t\bar{t}$ \cite{ATLAS:2020ccu,ATLAS:2022mlu} \\
        \swatch{blue}~$\ell$+\MET{}+jet \cite{ATLAS:2017luz} &
        \swatch{turquoise}~$\ell_1\ell_2$+\MET{}+jet \cite{ATLAS:2019ebv} &
        \swatch{green}~\MET{}+jet \cite{ATLAS:2017txd,ATLAS:2024vqf} &
        \swatch{indigo}~$\ell^+\ell^-$+\MET{}   \\
        \swatch{silver}~jets  &
        \swatch{orange}~$\ell^+\ell^-$+jet  &
        \swatch{skyblue}~$\ell_1\ell_2$+\MET{}  & 
        \swatch{magenta}~4$\ell$  \\ 
    \end{tabular}  
        \caption{Results from \contur, showing the dominant signatures in different regions of parameter space. 
        a) $M_\chi=100~\GeV$
        b) $M_\chi=10~\TeV$
        c,d) $\MHB = 200~\GeV$.
        In all cases $\gB = 0.15, \stB = 0.01$, except for figure d) with $\gB = 0.25$.
        The region between the black dashed lines is disfavoured at the 68\% confidence level, and
        between the black solid lines at the 95\% confidence level. The region between the red dotted
        lines is the region to which the HL-LHC is estimated to be sensitive (see text).
        d) Has the same configuration as c) but with $\gB = 0.25$.}
        \label{fig:conturScan1}             
\end{figure}

Fig.~\ref{fig:RivetPlots} demonstrates a subset of differential cross section measurements used as inputs to derive collider constraints with \contur.
The plots indicate experimentally observed data with black markers, the SM prediction as a red line and the BSM signal superimposed on the SM background as a blue line.
The ratio panel shows the agreement of the SM and SM+BSM prediction relative to the data, where the vertical axis is scaled to the 
total uncertainty to show agreement expressed as the number of standard deviations it is away from the data. These differential cross section measurements are chosen as
a representative subset of the processes that drive the 2D observed exclusions.
Note that no uncertainties are available for SM prediction for boosted $t\bar{t}$
pairs. The range of predictions presented in Refs.~\cite{ATLAS:2022mlu,ATLAS:2020ccu} shows a spread indicating a significant uncertainty; however this
spread is generally smaller than the experimental uncertainty, especially in the key $t\bar{t}$ mass measurement, and none produce a resonant bump as seen in our model.
In Fig.~\ref{fig:conturScan1} we show the dependence of the sensitivity on other parameters of the model. Figs.~\ref{fig:conturScan1}a, b and c use a coupling $\gB = 0.15$, which the \ZB search limits (Fig.~\ref{ZBbounds}) leave open. Fig.~\ref{fig:conturScan1}d uses $\gB = 0.25$, which is on the boundary of exclusion
but which may be allowed within the necessary approximations used in 
reinterpreting the searches in Fig.~\ref{ZBbounds}, particularly at low \MZB.

The existing measurements have little sensitivity (none, at the 95\% level for $\gB = 0.15$), and also exhibit only a weak dependence on \MHB and $M_\chi$ for this value of coupling. Note that actual exclusion exceeds the expected exclusion, because the SM predictions for $t\bar{t}$ production, which as mentioned have no SM uncertainties assigned, already lie above the data at high transverse momentum. Conversely, there is an expected sensitivity at low \Mchi,  which is not seen in the data; this comes from
the monojet measurements. In Ref.~\cite{ATLAS:2024vqf} it was shown that for these measurements to have comparable sensivity to the monojet searches, the ratio to
leptonic measurements must be used; however, at present only the cross section measurements are used.
\section{DARK MATTER CONSTRAINTS}
\label{Sec5}
The Majorana DM candidate in these theories can annihilate mainly into the following channels:
\begin{equation}
   \chi \chi \to  q \bar{q}, \ZB \ZB, \ZB \HB, \HB \HB, h h, h \HB, WW, ZZ. 
\end{equation}
Using the standard freeze-out calculation we can define the parameter space which is consistent with the cosmological bounds on the DM relic density. This DM candidate has been discussed in detail in Refs.~\cite{FileviezPerez:2019jju,FileviezPerez:2020oxn}. Here we revisit this study and use the new bounds from direct detection experiments.
\begin{figure}[h]
         \centering
    \begin{subfigure}[b]{0.9\textwidth}
         \centering
         \includegraphics[width=\textwidth]{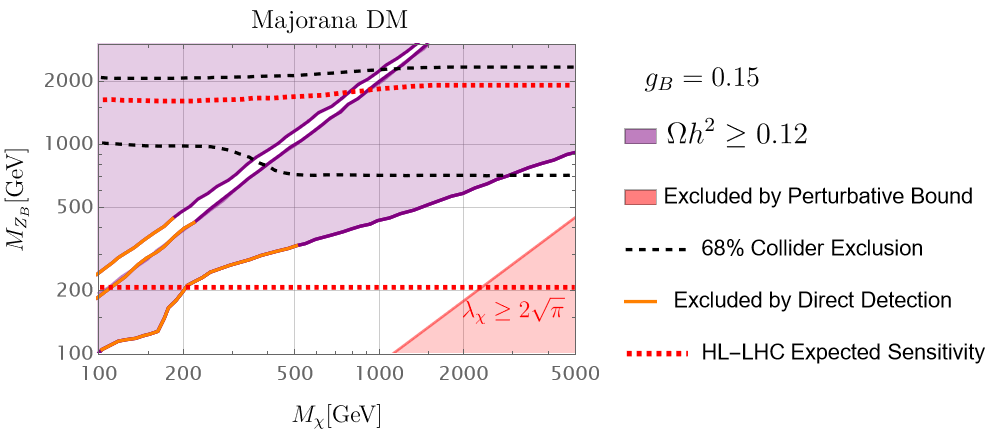}
         \caption{}
    \end{subfigure}  
         \begin{subfigure}[b]{0.9\textwidth}
         \centering
         \includegraphics[width=\textwidth]{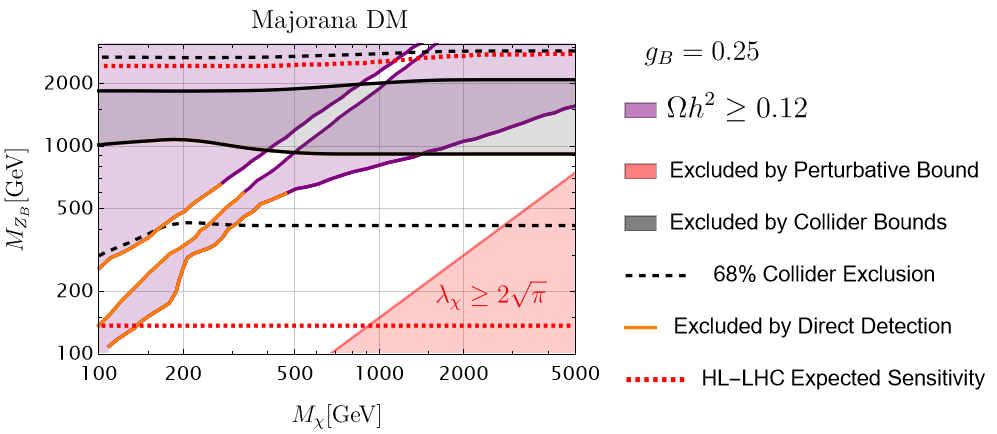}
         \caption{}
    \end{subfigure}  
        \caption{Correlation between relic density, direct detection, perturbative and collider bounds. Here the purple solid line satisfies the relic density constraints $\Omega_\chi h^2 = 0.1198 \pm0.0022 $~\cite{Planck:2018vyg} while the purple shaded region represents the parameter space when the produced relic density is greater than 0.12. The red-shaded region is excluded by the perturbative bound on the Yukawa coupling $\lambda_{\chi} \geq 2 \sqrt{\pi}$. Here we assumed $\MHB = 200$ GeV. The 68\% (dashed black line) and 95\% (solid black line) collider bounds from \contur are shown, as given in Fig.\ref{fig:conturScan1}c and Fig.\ref{fig:conturScan1}d for  $\gB = 0.15$ and $\gB = 0.25$, respectively. The region between the red dotted
        lines is the region to which the HL-LHC is estimated to be sensitive (see text). The orange line corresponds to the scenarios ruled out by direct detection bounds..
}
        \label{DMannhiall}
\end{figure}
In Fig.~\ref{DMannhiall} we show the allowed values for the DM and gauge boson mass according to these cosmological bounds, i.e. $\Omega h^2 \leq 0.12$ when $\gB=0.15$ (upper panel) and when $\gB = 0.25$ (lower panel). Notice that in this case one can satisfy the collider bounds in Fig.~\ref{ZBbounds} and~\ref{fig:conturScan10} when $\gB=0.15$, but in the case when $\gB = 0.25$ the collider bounds  exclude a fraction of the parameter space identified by the black-solid lines in the lower-panel. 
The allowed regions have the following features: a) the small allowed region around the line $\MZB \sim 2 M_\chi$ corresponds to the annihilation into two quarks via the \ZB resonance. In the other region, which is more generic, the gauge boson mass must be below the multi-TeV scale and one can find an upper bound on the symmetry-breaking scale. In this region the dominant annihilation channel is the $\chi\chi \to \HB \ZB$. The region in light-red is excluded when one imposes a perturbative bound on the Yukawa coupling $\lambda_\chi$. 
The line in orange is excluded by the direct detection bounds and the two dashed lines show the region that is excluded by collider bounds only at $68 \%$ confidence level when $\gB=0.15$ (upper panel). As one can appreciate from Fig.~\ref{DMannhiall} there is a large fraction of parameter space that is still unconstrained by collider or direct detection bounds even if \MZB and $M_\chi$ are very close to the electroweak-scale.

\begin{figure}[h]
    \centering
    \includegraphics[width=0.7\linewidth]{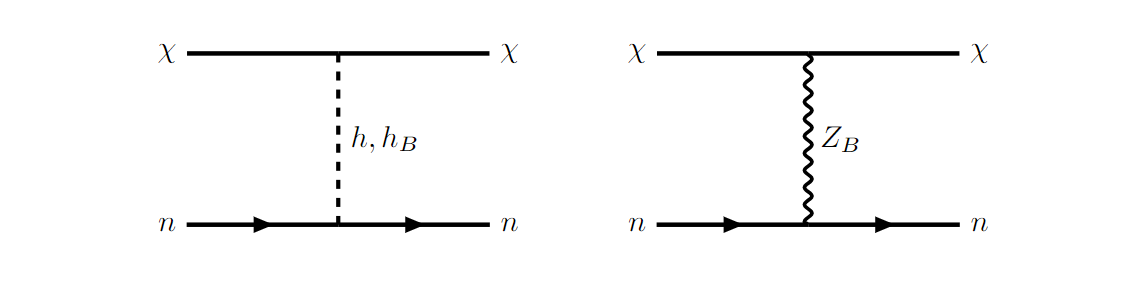}
    \caption{Feynman diagrams for DM-nucleon cross section.}
    \label{fig:feynmanDD}
\end{figure}

The DM candidate, $\chi$, can interact with quarks by exchanging the gauge boson \ZB and the Higgs bosons through the mixing angle \tB. The Feynman diagrams for these processes is shown in Fig~\ref{fig:feynmanDD}. The spin-independent cross section mediated by the physical Higgs bosons is suppressed by the mixing angle, and the cross-section can be written as 
\begin{equation}
    \sigma_{\chi N}(h_i)=\frac{72 G_F}{\sqrt{2} 4 \pi}\sin^2{\tB}\cos^2{\tB} \ M_n^4 \ \frac{\gB^2 M_{\chi}^2}{\MZB^2}\left(\frac{1}{M_h^2}-\frac{1}{\MHB^2}\right)^2 f_{N}^2.
\end{equation}
Here $f_N$ is the effective nucleon-nucleon-Higgs coupling~\cite{PhysRevLett.119.181803} and $M_n$ is the nucleon mass.
The spin-independent cross section mediated by the \ZB gauge boson is velocity suppressed and given by
  \begin{equation}
    \sigma_{\chi N}(\ZB)=\frac{27}{8\pi}\frac{\gB^4 M_n^2}{\MZB^4} v^2.
\end{equation}
Since this cross section is velocity suppressed one can satisfy the experimental bounds from DM direct detection even if the $\MZB$ is close to the electroweak scale.
\begin{figure}[t]   
  \begin{subfigure}[t]{0.6\textwidth}
    \centering
    \includegraphics[width=\textwidth]{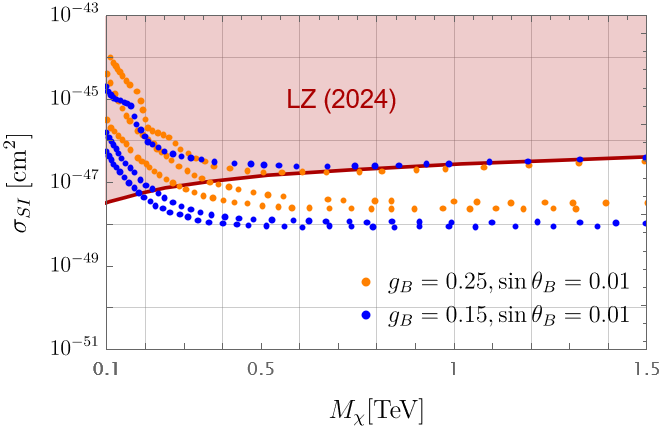}
  \end{subfigure}
       \caption{ Numerical values for the spin-independent cross-section for DM-nucleon scattering. The red-shaded region is excluded by the LZ (2024) experiment~\cite{LZ}. Here we used $\MHB$=200 GeV and $f_N$=0.3.~\cite{PhysRevLett.119.181803}.} 
       \label{fig:DMDirect1} 
     \end{figure}
     
In Fig.~\ref{fig:DMDirect1} we show the predictions for the spin-independent cross section, $\sigma_{SI} = \sigma_{\chi N}(\ZB)+ \sigma_{\chi N}(h_i) $ and the experimental bounds from the LZ collaboration. As one can appreciate, only in the region when the $M_\chi \sim 100-200$ GeV one can exclude some scenarios with $\gB=0.15$. In the case with $\gB=0.25$ one can exclude regions with the DM mass below $300$ GeV. In Fig.~\ref{fig:DMDirect1} we show only the scenarios represented by the blue ($\gB=0.15$ and $\sin \tB=0.01$) or orange ($\gB=0.25$ and $\sin \tB=0.01$) dots where one has the correct relic density, i.e. $\Omega_\chi h^2=0.12$. The structure of the allowed regions in Fig.~\ref{DMannhiall} thus gives three predictions
for the cross section for each value of $M_\chi$, where each prediction corresponds to a different value of \MZB. It can be seen that in this theory one can satisfy the dark matter direct detection constraints even when the dark matter and the leptophobic gauge boson masses are close to the electroweak scale.

\section{FUTURE PROSPECTS}
\label{Sec6}
%
\begin{figure}[h]
         \centering
    \begin{subfigure}[b]{0.45\textwidth}
         \centering
         \includegraphics[width=\textwidth]{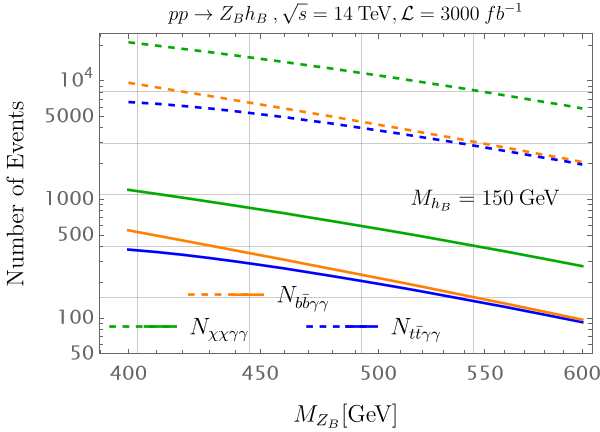}
         \caption{}
    \end{subfigure}  
         \begin{subfigure}[b]{0.45\textwidth}
         \centering
         \includegraphics[width=\textwidth]{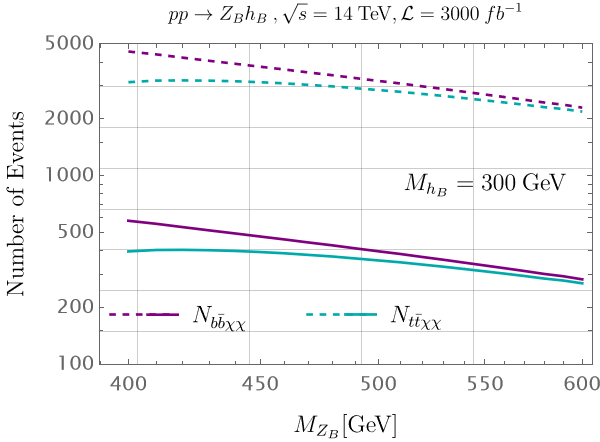}
         \caption{}
    \end{subfigure} 
        \caption{ (a) Number of events for different signatures as a function of \ZB mass when $\MHB = 150$ GeV. (b) same as (a), but now $\MHB=300$ GeV. The solid and dashed lines represent the number of events for $\gB =0.15$ and $\gB = 0.25$, respectively.  Here we assumed $M_{\chi}=100$ GeV , $\sqrt{s}=14$ TeV, the luminosity is $\mathcal{L}=3000$ $\text{fb}^{-1}$ and used MadGraph5~\cite{Alwall:2011uj} to compute the cross section for $p p \rightarrow \ZB \HB$.}
        \label{Nevents}
\end{figure}
It is striking that, as discussed in the previous section, a well-motivated, anomaly-free
theory containing a DM candidate can lie so close to the EW scale and yet be compatible with all current collider, cosmological and direct-detection limits. The phenomenology
suggests a discovery at the LHC is possible, and indeed potentially imminent. To illustrate this, we have estimated the sensitivity of the final HL-LHC dataset by scaling the experimental uncertainties of the measurements made at $\sqrt{s} = 13$~TeV, by the square root of the ratio of the current integrated luminosity used to the final HL-LHC sample - assumed to be \ab{3}.This naive estimate therefore assumes that i) the experimental systematic uncertainties scale with luminosity along with statistics ii) the phase space of the measurements is not extended despite the increase in statistics and beam energy (currently $\sqrt{s} = 13.6$~TeV) iii) no new final-state signatures are measured. The projected sensitivity at the 95\% confidence level under these assumptions is indicated by red dotted lines on Figs.~\ref{fig:conturScan10}, \ref{fig:conturScan1} and \ref{DMannhiall}. From this it is
clear that significant potential exists to probe this model at colliders in the short-to-medium term over much of the parameter space allowed by other constraints, although
this will require precise SM calculations, particularly of $t\bar{t}$ at high $p_T$.

Overall, the assumptions made in estimating this potential are likely to be pessimistic, so in Fig.~\ref{Nevents} we show the raw numbers of events expected from the HL-LHC for a selection of scenarios and final states, illustrating the experimental and theoretical challenge which may be met by improved analysis techniques and more precise SM predictions.
In Fig.~\ref{Nevents}a we show the number of events using the production mechanism $pp \to \ZB \HB$ when the Higgs mass is $\MHB=150$ GeV. We show the predictions for channels with two $b$-quarks and two photons ($N_{b\bar{b}\gamma \gamma}$), for channels with missing energy and two photons ($N_{\chi\chi\gamma \gamma}$), and for channels with two top-quarks and two photons ($N_{t\bar{t}\gamma \gamma}$). The solid and dashed lines represent the number of events for $\gB =0.15$ and $\gB = 0.25$, respectively.  Here we assumed $M_{\chi}=100$ GeV, $\sqrt{s}=14$ TeV and a total integrated luminosity of ${\mathcal{L}=3000\,\text{fb}^{-1}}$. In Fig.~\ref{Nevents}b we show the predictions for the number of events when $\MHB=300$ GeV for the channels with missing energy and two heavy quarks.
\section{SUMMARY}
\label{Sec7}
We have discussed a general class of theories where a DM candidate is predicted from anomaly cancellation. 
A specific gauge theory, in which the new symmetry is local baryon number, is investigated in detail. In this context the stability of DM is a natural consequence of the spontaneous breaking of the new gauge symmetry. 
We discussed the case where the DM candidate is a Majorana fermion. 
The new physical Higgs, \HB, responsible for symmetry breaking in the theory, has very interesting properties, because its decay has a large branching two photons in the low mass region and, if kinematically allowed, to DM in the intermediate mass region. 

The LHC constraints for this model are evaluated, delineating the allowed masses for the new gauge boson \ZB associated with baryon number gauge symmetry. In this case, \ZB is leptophobic, and decays into quarks or DM, since the exotic fermions needed for anomaly cancellation are assumed to be heavy. The principle sensitivity in LHC measurements and searches comes from resonant quark production, especially $t\bar{t}$ when \MZB is large enough, and \MET, if \MZB is low and $2\Mchi < \MZB$. When these two
principle channels are kinematically disallowed, production of \HB in association with \ZB, followed by \HB decay to photons, can play a significant role.

The relic density constraints are evaluated, showing the allowed parameter space and the scenarios that can satisfy direct detection constraints. We have shown the overlap between the collider constraints, the relic density constraints and the direct detection bounds. There is a significant open parameter region at low \ZB masses ($\MZB \lesssim 500~\GeV$) where the new Higgs is also light, which should be well within reach of future (HL-)LHC analyses of the signatures discussed. This study tells us that one can hope to discover a simple theory for DM in the near future.

{\small{{\textit{Acknowledgments}}: The work of H.D. and P.F.P. is supported by the U.S. Department of Energy, Office of Science, Office of High Energy Physics, under Award Number DE-SC0024160. This work made use of the High Performance Computing Resource in the Core Facility for Advanced Research Computing at Case Western Reserve University, and of the HEP computing cluster at UCL supported by STFC. Y. Y. thanks the Elizabeth Spreadbury fund for his PhD studentship. We thank Joe Egan for useful discussions. }}
\appendix
\section{FEYNMAN RULES}
\begin{align}
    q \bar{q} \ZB  &: \hspace{0.5cm}  i\frac{\gB}{3} \gamma^\mu, \\
    \chi \chi \ZB  &: \hspace{0.5cm}   i\frac{3}{2}\gB \gamma^\mu \gamma^5,\\
    \chi \chi \HB &: \hspace{0.5cm}   - i \frac{3 \gB M_{\chi}}{\MZB} \cos \tB,\\
     \chi \chi h &: \hspace{0.5cm}    i \frac{3 \gB M_{\chi}}{\MZB} \sin \tB, \\
     \HB \ZB \ZB &: \hspace{0.5cm}  6 i\hspace{0.1 em} \cos \tB \hspace{0.1 em} \gB\hspace{0.1 em} \MZB \ g^{\mu \nu},\\
     h \ZB \ZB &: \hspace{0.5cm} - 6 i\hspace{0.1 em} \sin \tB \hspace{0.1 em} \gB\hspace{0.1 em} \MZB \ g^{\mu \nu}, \\
     h Z Z & : \hspace{0.5 cm} 2i  \frac{M_{Z}^2}{v_0} \cos{\tB} \ g^{\mu \nu},\\
      \HB Z Z & : \hspace{0.5 cm}  2i \frac{M_{Z}^2}{v_0} \sin{\tB} \ g^{\mu \nu}, \\
      h W W & : \hspace{0.5 cm} 2i   \frac{M_{W}^2}{v_0} \cos{\tB} \ g^{\mu \nu},  \\
      \HB W W & : \hspace{0.5 cm}  2i   \frac{M_{W}^2}{v_0} \sin{\tB} \ g^{\mu \nu},      \\ 
h h \HB &: i [ - 6  \lambda_H v_0 \cos^2 \tB \sin \tB- 2 \lambda_B  \cos \tB \sin^2 \tB\frac{\MZB}{\gB}  \nonumber+ \\
& \lambda_{HB} (-  \cos^3 \tB\frac{\MZB}{3 \gB}+ 2 v_0 \cos^2 \tB \sin \tB + 2 \cos \tB \sin^2 \tB \frac{\MZB}{3 \gB} - v_0 \sin^3 \tB ) ].
\end{align}

\bibliography{refs}

\end{document}